# OPTIMIZATION AND EVALUATION OF A MULTIMEDIA STREAMING SERVICE ON HYBRID TELCO CLOUD

Trong Duong Quoc[1] , Heiko Perkuhn[2], Daniel Catrein[2], Uwe Naumann[3] and Toni Anwar[4]

[1]The Sirindhorn International Thai German Graduate School of Engineering (TGGS)
King Mongkut's University of Technology North Bangkok (KMUTNB)
trongdq@gmail.com
[2]Ericsson Gmb, Ericsson Allee, Herzogenrath, Germany
heiko.perkuhn@ericsson.com
daniel.catrein@ericsson.com
[3]STCE, RWTH Aachen University, Seffenter Weg, Aachen, Germany
naumann@stce.rwth-aachen.de
[4]Universiti Teknologi Malaysia, Malaysia

## *ABSTRACT*

*With recent developments in cloud computing, a paradigm shift from rather static deployment of resources to more dynamic, on-demand practices means more flexibility and better utilization of resources. This demands new ways to efficiently configure networks.*

*In this paper, we will characterize a class of competitive cloud services that telecom operators could provide based on the characteristics of telecom infrastructure through an applicable streaming service architecture. Then, we will model this architecture as a cost-based mathematic model. This model provides a tool to evaluate and compare the cost of software services for different telecom network topologies and deployment strategies. Additionally, with each topology it acts as a means to characterize the deployment solution that yields the lowest resource usage over the entire network. These applications are illustrated through numerical analysis. Finally, a proof-of-concept prototype is deployed to shows dynamic properties of the service in the architecture and the model above.*

## 1. INTRODUCTION

Cloud computing has emerged as one of the hottest concepts in Information and Communication Technology (ICT) today. The big possible savings promised by virtualization and on-demand resource usage also attract the telecom industry. Services in today's networks offer typically much more capacity than needed on average. The dimensioning is determined by the expected peak load, which means that most of the deployed hardware is underutilized most of the time. By deploying services in Clouds inside the networks operators could make better use of the underlying hardware. This would increase utilization, flexibility and reduce the costs of their own operations [8]. Another advantage would be the possibility to exploit new business opportunities [8]. By increasing the flexibility of internal resources, cloud computing allows operators to use excess capacity for additional services and applications.

Under these circumstances, new solutions for planning, dimensioning and deploying resources in the network are needed. The approach introduced in this paper serves as a rather general foundation of optimization solutions - also for future services and network topologies. It is based on mathematical methods for optimizing a cost function under constraints. In order to be able to formulate the function, a model taking some real-world assumptions into account was created. The aim of this model is to provide an analytical tool that can be used to explore different aspects of the network in terms of cost-reduction.

To do that, the first contribution of this paper is to characterize a class of competitive cloud services that telecom operators could provide. These services are based on the cloud computing trend and the telecom topology. Specifically, we propose an applicable service architecture in which a telecom network (consists of primary sites and secondary sites) and third party data centers (represented by a public cloud, i.e. Amazon EC2) are taken into account in the deployment of application servers. Therefore, the service could exploit the local presence of secondary sites to enhance the end-user's experience through edge delivery methods and utilize public clouds to handle unpredictable peak loads. A multimedia streaming service is chosen to represent the targeted service group because of its high bandwidth requirement and drastic increase to the largest share in mobile networks.

The second contribution is a cost-based mathematic model. Based on the service architecture in the first contribution, we will model it mathematically. The network topology is first abstracted as nodes (public cloud, primary and secondary sites) and links (backbone links and links to external networks). Then the total cost function for the resource usage of a service is formulated including the server operation cost in data centers (internal and external) and link cost for data transfer. Then by means of minimizing the cost function, the model will determine the necessary number of streaming servers, server placement and request-routing scheme such that the service can serve all the client requests sufficiently and in a way that minimizes the cost for operators. Using the proposed model, operators can compare and evaluate the cost of software services for different telecom network topologies and deployment strategies as well as characterize the optimal deployment solution. These useful applications of the model are demonstrated by analytical and numerical results.

The final contribution of this paper is a proof-of-concept prototype of a live streaming service. Signals from a satellite are encoded into a stream which will be sent to streaming servers to serve client requests. The prototype shows the fast and easy deployment of media streaming servers in different clouds. With its architecture, the load on the head-end is constant while the number of streaming servers can be scaled up according to the demand caused by clients.

The rest of the paper is organized as follows: Section 2 describes the background on the mobile network topology and Telco cloud; Section 3 depicts the service architecture; Section 4 introduces the mathematical model and related assumptions; in Section 5 the different solution approaches to the optimization problem are evaluated; the proof-of-concept-prototype is described in Section 6; and Section 7 concludes the paper.

## 2. BACKGROUND ON TELCO CLOUD

### 2.1. Telecom network topology and site

The following is based on the work in [2], where the authors introduce a network topology and site model of typical mobile networks of today, comprising geographically separate primary and secondary sites. A typical mobile network consists of about 3 to 5 primary sites and 5 to 15 secondary sites.

Figure 1 depicts schematically the structure of such a network. Primary sites host the main functional elements of a mobile network. This includes service nodes, servers and external network connectivity. They can be compared with data centers and are interconnected via a backbone.

Secondary sites serve as concentration and distribution sites. They are typically much smaller than primary sites and host network equipment like media gateways (MGw), radio network controllers (RNC) or base station controllers (BSC). Secondary sites are also suited to host server type nodes, such as caches or media streaming servers. However, if these exist in today's architectures, they are statically deployed accruing to long term capacity and demand predications.

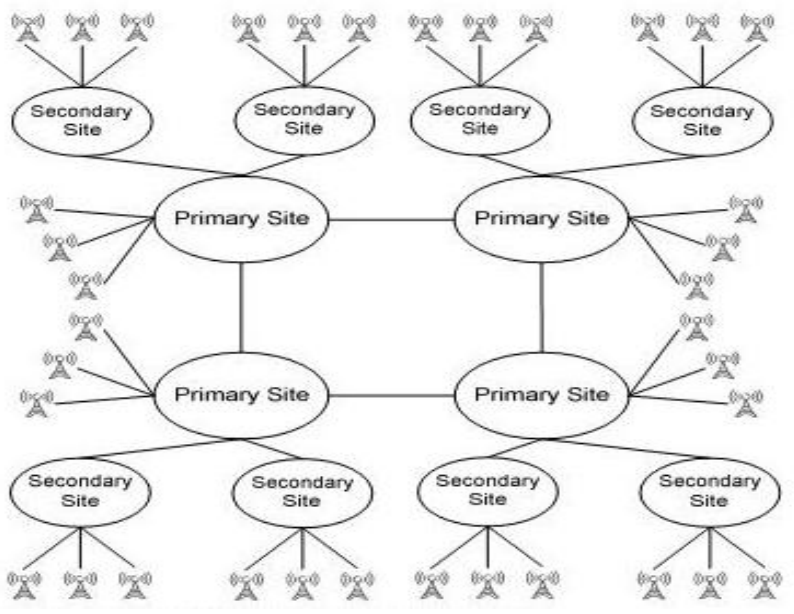

Figure 1. Schematic structure of a mobile network

Typically, secondary sites are connected to two primary sites (dual-homed) for redundancy reasons. Note that Figure 1 only shows a single connection to ease readability. Secondary sites connect to external networks and other sites via the primary sites they are connected to. However, with the increase in mobile data traffic, e.g., due to the evolution of HSPA and the rise of LTE, secondary sites also tend to have a direct connection to external networks, i.e., the Internet.

## 2.2. Telco Cloud

Cloud Computing is about the paradigm of offering computing resources as a Service in a flexible way. Examples for these resources are:

- Infrastructure-as-a-service: computing power, storage, networking
- Platform-as-a-service: e.g. application servers, queuing or load balancing
- Software-as-a-service: web mail, SAP services etc.

Cloud computing technology is finding its way also into telecom networks because of its increased flexibility, its cost efficiency and its potential to enable new service offerings. The virtualization of network services may lead to a much more efficient utilization of resources as well as a reduction of energy costs [3]. In this context ``private clouds" are usually mentioned, meaning the application of cloud technology to the internal operations. When it comes to offerings of cloud services from telecom operators towards third parties, there are currently only vague predictions possible [4]. As IT companies (like Amazon, Google, Microsoft or HP), equipment vendors (ALU, Ericsson etc.) and operators (Deutsche Telekom, KPN, Vodafone) all have different business models, they all have their own flavour and characteristics of cloud computing. In this paper we concentrate mainly on aspects concerning cost-efficient offering of media services in telecom networks.

In this paper, the notion of a ``Telco Cloud" refers to a telecom network in which IaaS resources such as virtual machines can be spawned as a Cloud service inside the network, e.g. in primary sites, in secondary sites or even at base stations. Figure 2 shows exemplarily such a Telco cloud, which consist of numerous smaller clouds distributed geographically.

Note that costs and effort wise, smaller sites with a small number of resources will be less efficient than very large sites. However, this is only true as long as we are only considering cost for processing and not the costs for networking. For high bandwidth applications, the overall picture may look very different as analyzed in this paper.

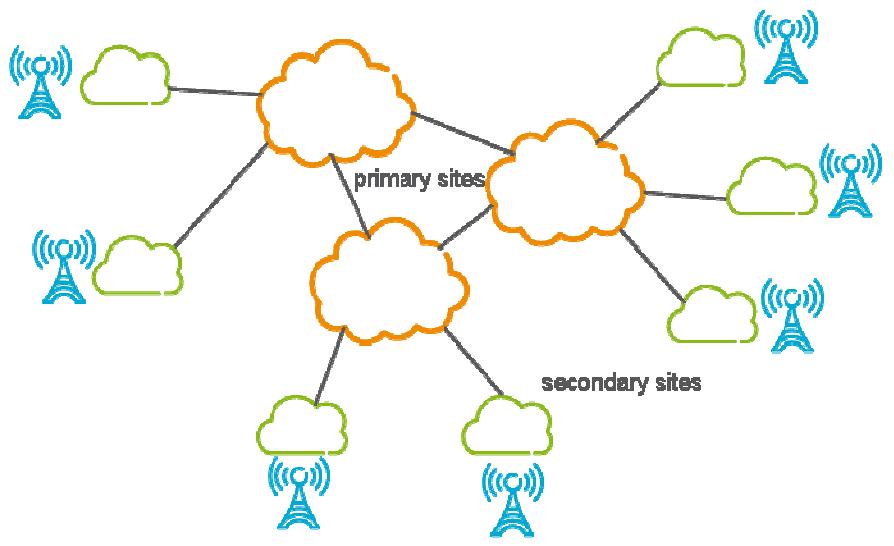

Figure 2. Telco Cloud with geographically distributed smaller clouds

## 3. SERVICE ARCHITECTURE

### 3.1. Architecture

In the near future, they could engage in cloud computing in different directions. In this section, we will characterize a competitive class of cloud services that operators could provide (see [21] for more details. These services are based on the cloud computing trend and especially the characteristics of telecom infrastructure (section II_A). To characterize these services, first of all we address a class of applications and services that requires low latency and high bandwidth which are difficult to perform well and deliver the expected quality of experience (QoE). Particularly, we choose a multimedia streaming service to represent the targeted service group because of two reasons. First, multimedia streaming, with its distinguishing characteristics, requires support on delay guarantees, bandwidth reservation, and flexible error control, which makes it challenging. Second, multimedia content has grown substantially over the past years and gained the largest share of the data traffic in mobile networks [10].

Standing on this point, we propose an applicable architecture for a multimedia streaming service on the Hybrid Telco Cloud as in Figure 3. In this architecture, the telecom network (consists of primary sites and secondary sites) as well as third party data centers (represented by public cloud, i.e. Amazon EC2) are taken into account for deployment of application servers (which are the streaming servers in the rest of the paper). Therefore, the service on one hand would exploit strengths of the operators' mobile networks (local presence of secondary sites) to enhance end user experience through edge delivery methods. On the other hand, it can be scaled dynamically, even to public clouds, to meet the changing demand of clients without worrying about unpredictable peak loads. While the first requirement is resolved quite well (but not yet mature), especially with the emergence of CDN and edge delivery methods, the second one is somehow more open, particularly in handling peak load ([6], section II_A).

In this architecture, streaming servers are predestined for being deployed as virtual resources in view of the fact that the demand may be of a spiky nature and is often not well predictable. Static deployments can easily face underutilization, which means sunk cost in infrastructure, or they are under-provisioned, which drives users away. Therefore, we introduce in the heart of the architecture a common management framework so-called cloud manager. It provides a single interface to applications, such as the streaming server, in order to request new resources and release those that are idle. The cloud manager allocates the resources by creating and configuring Virtual Machine (VM) images in different clouds (private and public ones) to meet the needs of the proposed streaming cloud service. Thus, the best balance between low cost and provisioning during demand peaks can be found.

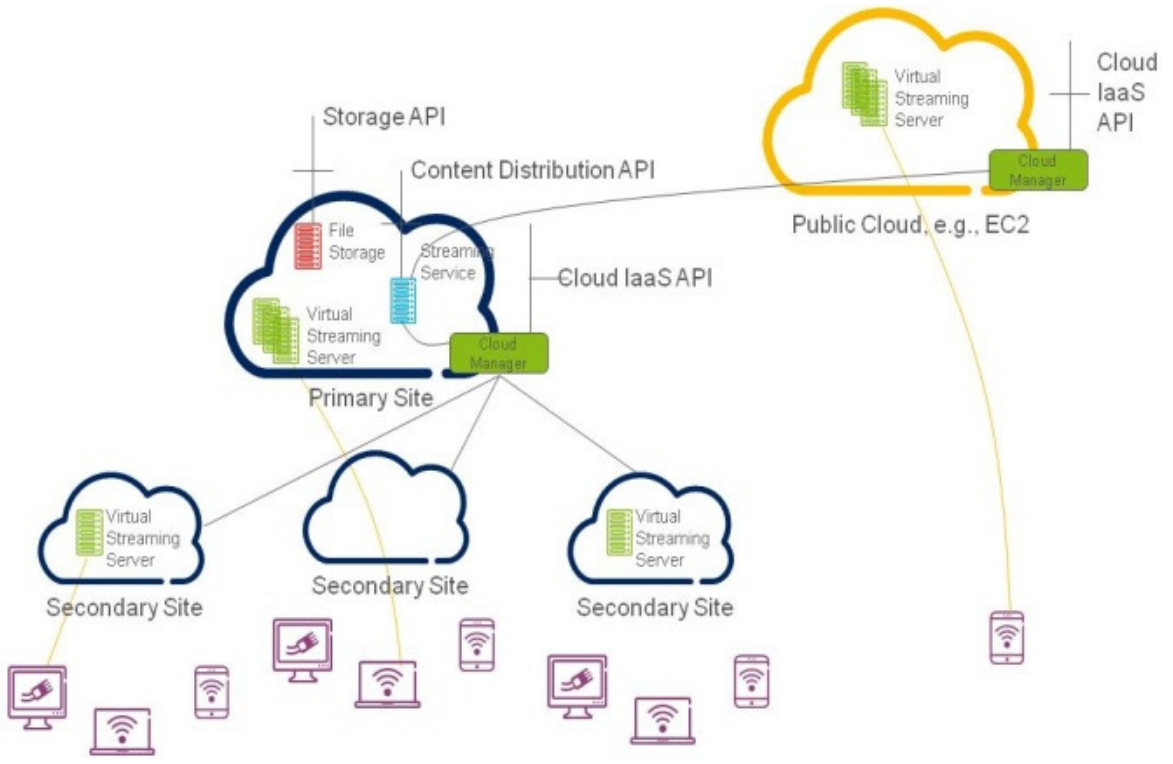

Figure 3. Architecture of streaming service on Hybrid Telco Cloud [21]

**The Cloud APIs**: Besides, there are also three Cloud APIs presented in the service architecture: The Cloud IaaS APIs, which allow cloud infrastructure to be added, reconfigured, or removed; The storage APIs allows content providers worldwide to store their media content in their own way; The content distribution APIs allow content providers to implement their strategy in delivering content through system's network. These 3 types of extensive API above allow third parties to participate and deliver their multimedia content. The variety of the content will determine the success level of streaming service.

**Working mechanism**: Based on the demand, the cloud manager will decide where and when to deploy the virtual machine (VM) instances. When a client streaming request is sent to the cloud manager through a website or portal, the cloud manager will determine the location of the request and then redirect that request to a streaming server in the respective network that is close to the client who sent the request.

In order to control the number of instances dynamically based on request and demand, the system will determine one ``warning threshold". At the beginning when the client sends the first request to the cloud manager, it will create a new VM instance to serve this request. If the number of requests continues to increase and exceeds the ``warning threshold", the cloud manager will start new instances and serves these new exceeded requests from these new instances.

Vice versa, at certain times when the number of requests is decreased so that they need fewer servers than the current available servers, there will appear one or several servers which are ``idle" - doesn't serve any client. These servers therefore would be terminated to both release resources for another application and reduce operation cost. In this way, the system can be scaled dynamically based on the demand.

## 3.2. Advantages and Evaluation

The proposed streaming cloud service can guarantee two requirements: ensure good user experience while scale up easily to handle workload efficiently.

Firstly, the service has clearly exploited telecom operators' strengths to ensure good user experience. In the architecture, the streaming server is brought out of a centralized data center to the far away sites which are closer to the user. This is here where the strengths of operator show up. With the widespread network of telecom sites, i.e. 15-20 sites per country [2], operators can provide global reach with domestic and international streaming points of presence. Meanwhile, as one of the leading cloud service providers in non-telecom industries, Amazon CloudFront provides only 17 edge locations worldwide as of December, 2010 [16]. The short distance between operators and users would reduce round-trip time, package loss and avoid congestion. As a result, multimedia streams are delivered with high quality and thereby enhance user experience. Besides, it is also easy to address relevant concern, i.e. trust or security...

Secondly, with the described dynamical scaling characteristic, the service allows operators to handle workload efficiently. They can exploit resources (underlying hardware) the most because each service will get only required resources and leave the rest for other services. Therefore there isn't any resource that will be left underutilized. Additionally, by automation of deployment and configuration, operators can reduce time and effort for operating and maintaining applications also. Furthermore, streaming cloud services allow operators to avoid the over provisioning issue to handle peak load. They don't need to worry if the peak load is out of their capacity because they can rent resources from public clouds. There are no long-term contracts additional charges; they simply pay for what they use.

Thus, together the Telco Clouds and Cloud manager make it possible to manage the demand for computing and network resources instantaneously, and to meet changing service needs more quickly and efficiently than today. However, the question of how to utilize cloud technologies in the best way, is however yet to be answered. As seen in the previous section, telecom networks have a certain topology. In order to exploit the characteristics of telecom networks in the context of cloud computing in the best way, a mathematical model describing the network is introduced in the next section.

# 4. MATHEMATICAL OPTIMIZATION MODEL

## 4.1. Purpose and Assumptions

In this section, the architecture in Section 3 will be modelled as a cost-based mathematic model to analyze and evaluate different aspects of the cloud-based telecom network. In order to do that, a total cost function for resource usage of the service is formulated. This total cost includes server operation costs in data centers (internal and external) and link costs for data transferring. Then by means of minimizing the cost function, it is possible to determine the necessary number of streaming servers, server placement and request-routing scheme such that the service can serve all the client requests sufficiently. For modelling, several assumptions on the mobile network topology are made.

**Assumption 1**: Typically, secondary sites are connected to two primary sites (dual-homed) for redundancy reasons. However, in this paper we assume that one secondary site connects to one primary site only for simplicity, even though the model could be extended to cover also the other case.

**Assumption 2**: We also assume that secondary sites connect to an external network and other sites via the primary site they are connected to. This reflects the current status of mobile networks on the market. However, if the secondary site has the direct connection to an external network in the future, the model can be changed accordingly (see [14]).

**Assumption 3:** The order of redirection: because primary sites are typically much bigger than secondary sites, we assume that requests from clients connected to secondary site could be redirected to a primary site or public cloud but client requests to primary sites will not be redirected to secondary sites. In addition, in order to move peak loads to a public cloud, we also assume that requests to primary sites could be redirected to a public cloud.

## 4.2. Optimization Problem

We can start modelling by stating the optimization problem mathematically as following (illustrated in Figure 4):

The operator has $P_{num}$ number of primary sites and every $i^{th}$ primary site will connect with several secondary sites, *SAP[i]*. At a given moment, there are a number of clients connecting to each site ($C_p[i]$ clients for $i^{th}$ primary site and $C_s[i,j]$ clients for its $j^{th}$ secondary site) and requesting for streaming service.

The operator needs to deploy streaming servers into every site, including public clouds to serve client requests. Each server in a public cloud costs $P_a$ cents/hour. Each server in the $i^{th}$ primary site costs $P_p[i]$ cents/hour. Each server in the $j^{th}$ secondary site of the $i^{th}$ primary site costs $P_s[i,j]$ cents/hour.

Clients who connect to a secondary site can be served from a primary site or public cloud (redirection). In this case, the multimedia streams are sent through the link between the public cloud and a primary site, which takes $L_a$ cents/hour, and/or the link between a primary site and another primary site, which takes $L_{pp}$ cents/hour, and/or a the link between a primary site and a secondary site, which takes $L_s$ cents/hour.

Therefore, the operator wants to determine how many servers should they deployed at each site and what strategy be used as serving scheme so that client requests can sufficiently be served while it is still ensured that the cost is minimized.

## 4.3. Modelling

The network will be considered in a snapshot of an hour. First we describe the parameters and variables used. Then, we will state objective function and constrains. Afterwards, we provide detailed arguments for each of the terms in the objective function and constraints.

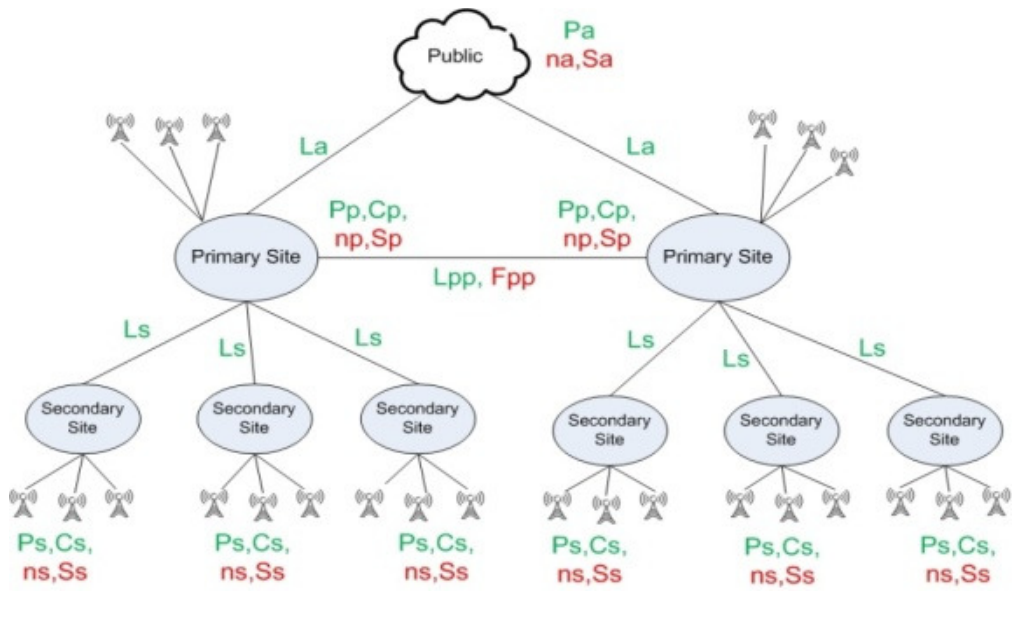

Figure 4. The basic model with all parameters and variables

Parameters:

| | |
|---|---|
| $P_{num}$ | : number of primary site |
| *SAP[i]* | : number of secondary site at i^th primary site |
| $U$ | : capability (number of users) per server |
| $P_a$ | : operation cost for one server on public cloud |
| $P_p[i]$ | : operation cost for one server on i^th primary site |
| $P_s[i,j]$ | : operation cost for one server on j^th secondary site |
| $C_p[i]$ | : number of clients connect to i^th primary site |
| $C_s[i,j]$ | : number of clients connect to j^th secondary site |
| $L_a$ | : link cost per client for the link btw public cloud and primary site |
| $L_{pp}[i,j]$ | : link cost per client for the link between pri. site i^th and j^th, i,j = 1..$P_{num}$, i <>j |
| $L_{sp}[i,j]$ | : link cost per client for the link btw pri. site i^th and sec. site j^th, i= 1..$_{Snum}$, j=1..$P_{num}$ |

Variables:

| | |
|---|---|
| $n_a$ | : number of servers on the public cloud |
| $n_p[i]$ | : number of servers on the i^th primary site |
| $n_s[i,j]$ | : number of servers on the j^th secondary site of i^th primary site |
| $S_a[i]$ | : number of clients served from public cloud |
| $S_p[i]$ | : number of clients served from i^th primary site |
| $S_s[i,j]$ | : number of clients served from j^th secondary site |
| $F_{pp}[i,j]$ | : number of clients redirected from i^th pri. to j^th pri. ,i,j=1..$P_{num}$, i <>j |

Then, we have the objective function with constraints:

$$\underset{n_a, n_p[i], n_s[i,j], S_a[i], S_p[i], S_s[i,j]}{\text{minimize}} \quad \text{Total Cost} = \underbrace{n_a \times P_a}_{A} + \underbrace{\sum_{i=1}^{Pnum} n_p[i] \times P_p[i]}_{B} + \underbrace{\sum_{i=1}^{Pnum} \sum_{j=1}^{SAP[i]} n_s[i,j] \times P_s[i,j]}_{C}$$

$$+ \underbrace{\sum_{i=1}^{Pnum} \sum_{j=1}^{SAP[i]} \left(C_s[i,j] - S_s[i,j]\right) \times L_s[i,j]}_{D}$$

$$+ \underbrace{\sum_{i=1}^{Pnum} S_a[i] \times L_a[i]}_{E} + \underbrace{\sum_{i=1}^{Pnum} \sum_{j=1, j \neq i}^{SAP[i]} F_{pp}[i,j] \times L_{pp}[i,j]}_{F}$$

Such that:

(C1): $\sum_{i=1}^{Pnum} S_a[i] \le n_a \times U$

(C2): $S_p[i] \le n_p[i] \times U$

(C3): $S_s[i,j] \le n_s[i,j] \times U$

(C4): $C_p[i] + \sum_{j=1}^{SAP[i]} C_s[i,j] + \sum_{j=1, j\neq i}^{Pnum} F_{pp}[j,i] = S_a[i] + S_p[i] + \sum_{j=1}^{SAP[i]} S_s[i,j] + \sum_{j=1, j\neq i}^{Pnum} F_{pp}[i,j]$

(C5): $C_s[i,j] \geq S_s[i,j]$

(C6): $n_a, n_p[i], n_s[i,j], S_a, S_p[i], S_s[i,j] \geq 0$, and $\in$ □

(C7): $n_p[i] \leq 800$ and $n_s[i,j] \leq 100$

**Formulating The Total Cost Function**

The total cost function consists of the total server operation cost and the total link cost.

The total server operation cost is the sum of the server operation cost at each site. This server operation cost at each site is calculated by multiplying the number of streaming servers running in that site with the cost to run each server in that site.

- There are $n_a$ servers on the public cloud and each server costs $P_a$ cents/hour, therefore the server operation cost on the public cloud is: $A = n_a \times P_a$.
- Similarly, there are $n_p[i]$ servers at the $i^{th}$ primary site and each server costs $P_p[i]$ cents/hour, so the operation cost at the $i^{th}$ primary site is: $n_p[i] \times P_p[i]$. Thus, the server operation cost of all primary sites is: $B = \sum_{i=1}^{Pnum} n_p[i] \times P_p[i]$.
- At the $j^{th}$ secondary site of the $i^{th}$ primary site, there are $n_s[i,j]$ servers and each server costs $P_s[i,j]$ cents/h, so the server operation cost of that secondary site is: $n_s[i,j] \times P_s[i,j]$. Therefore, the server operation cost of all secondary sites is: $C = \sum_{i=1}^{Pnum} \sum_{j=1}^{SAP[i]} n_s[i,j] \times P_s[i,j]$.

By aggregating all the operation costs at each site, we get the formula for the total server operation cost: **Total Server Operation Cost = A + B + C**.

The total link cost is the sum of the link costs at each link. The link cost at each link is calculated by multiplying the number of streams sent over that link (in the case of redirection) with the price for one stream in one hour.

- With the link between the $i^{th}$ primary site and its $j^{th}$ secondary site, the number of streams sent over it is the number of redirections from secondary site to primary site, or the difference between the number of clients connected to a secondary site ($C_s$) and the number of clients served from the secondary site ($S_s$): $C_s[i,j] - S_s[i,j]$. The price for one stream on this link is: $L_s[i,j]$ cents/hour. Thus, the link cost at this link is: $(C_s[i,j] - S_s[i,j]) \times L_s[i,j]$. Therefore, the link cost for all links of this kind: $D = \sum_{i=1}^{Pnum} \sum_{j=1}^{SAP[i]} (C_s[i,j] - S_s[i,j]) \times L_s[i,j]$
- With the link between the $i^{th}$ primary site and a public cloud (external network), the number of streams sent over it is the number of clients served from public cloud $S_a[i]$.

The price for one stream on this link is: $L_a[i]$ cents/hour. Thus, the link cost at this link is: $S_a[i] \times L_a[i]$. Therefore, the link cost for all links of this kind: $\sum_{i=1}^{Pnum} S_a[i] \times L_a[i]$

- With the link between the $i^{th}$ primary site and the $j^{th}$ primary site, the number of streams sent over it is the variable $F_{pp}[i,j]$. The price for one stream on this link is: $L_{pp}[i,j]$ cents/hour. Thus, the link cost at this link is: $F_{pp}[i, j] \times L_{pp}[i, j]$. Therefore, the link cost for all links of this kind: $F = \sum_{i=1}^{Pnum} \sum_{j=1, j \neq i}^{SAP[i]} F_{pp}[i, j] \times L_{pp}[i, j]$

By aggregating all the link costs for each link, we get the formula for the total link cost: **Total Link Cost = D + E + F.**

By sum of total link cost and total operation cost, we get the total cost function.

## Formulating Constraints

In the model above, there are 8 constraints (C1-C7) that variables have to satisfy.

The constraints C1, C2, C3 guarantee that the number of deployed streaming servers at each site is big enough to serve all the requests assigned to it. This is expressed by an inequality with the left side being the number of clients served by the telecom site, and the right side is the capacity provided by all deployed servers at that site.

- Constraint C1: The left side of C1 is the number of all clients served from a public cloud. Whereas on the right side of C1 we have the product of number of servers on the public cloud ($n_a$) and the capacity of one server ($U$ clients). This product is the number of clients that all deployed servers in the public cloud could serve. Therefore, Constraint C1 guarantees that we will deploy enough servers on the public cloud to serve all clients redirected to it.
- Constraint C2: The left side of C2 is the number of clients served from the $i^{th}$ primary site. Whereas on the right side of C2 we have the product of number of servers at that primary site ($n_p[i]$) and the capacity of one server ($U$ clients). This product is the number of clients that all deployed servers at that primary site could serve. Thus, Constraint C2 means that we will deploy enough servers at each primary site to serve all clients connected and/or redirected to that primary site.
- Constraint C3: Similarly, Constraint C3 ensures that we will deploy enough servers at each secondary site to serve all clients assigned to be served from that secondary site.

The constraint C4 ensures that all client requests are served, so that there isn't any client request left. This constraint applies for each $i^{th}$ primary site, $i = 1..P_{num}$ ($P_{num}$: number of primary sites).

- At the $i^{th}$ primary site, there are $C_p[i]$ clients requesting streams and $SAP[i]$ secondary sites connected to it. With each of these secondary sites, there are $C_s[i,j]$ clients requesting streams ($j=1..SAP[i]$). Additionally, there are $\sum_{j=1, j \neq i}^{Pnum} F_{pp}[j,i]$ requests redirected from other primary sites as well. Therefore, the total number of client requests for this $i^{th}$ branch (including the $i^{th}$ primary site and all of its secondary sites) is equal to the left side of C4.

- All of these client requests for this branch could be served from a public cloud ($S_a[i]$ requests), the $i^{th}$ primary site ($S_p[i]$ requests) or the $j^{th}$ secondary site of the $i^{th}$ primary site ($S_s[i,j]$ requests). They could also be redirected to other primary sites $\sum_{j=1, j \neq i}^{Pnum} F_{pp}[i, j]$. Thus, the right side of C4 is the total number of clients served in this $i^{th}$ branch.

After all, by the equality, constraint C4 guarantees that all clients connected to this branch are served from this branch as well.

The Constraint C5 indicates the fact that the number of clients connected to each secondary site ($C_s$[i,j]) is always greater or equal to the number of clients served from that secondary site ($S_s$[,j]). This means that there is only redirection from secondary site to primary site and there isn't any request from a primary site redirected to any secondary site as described in assumption 3 in the previous section.

The constraint C6 is rather straight forward. It describes that all the variables $n_a$, $n_p[i]$, $n_s[i,j]$, $S_a[i]$, $S_p[i]$, $S_s[i,j]$ are integer and are not negative numbers. Finally, the constraints C7 present the capacity of a primary site (800 servers) and a secondary site (100 servers) that will be explained in Section 4.5.

## 4.4. Optimization Solving Process

To optimize, we choose a software package including AMPL [11] modelling language and LPSolve solver [19]. AMPL is chosen because the similarity of its syntax to mathematic notation allows us to describe optimization problems very concisely while retaining the readability and the understandability of the problems. It also allows us to choose the external solver. Since our optimization problem is an integer programming problem [12], we choose the powerful open-source linear solver: LPSolve.

The solving process is presented in Figure 5. Firstly, the mathematic model in Section 4.3 is converted to an AMPL model using the AMPL modelling language. After that, the AMPL data file is created with input data for all the parameters in the AMPL model. Then the LPSolver, an optimization solver, is selected and downloaded. Finally, the AMPL script will call the AMPL model, AMPL data file and solver together to print out the optimal total cost with the respective values of variables (see [14] for more details of AMPL model, data file and script).

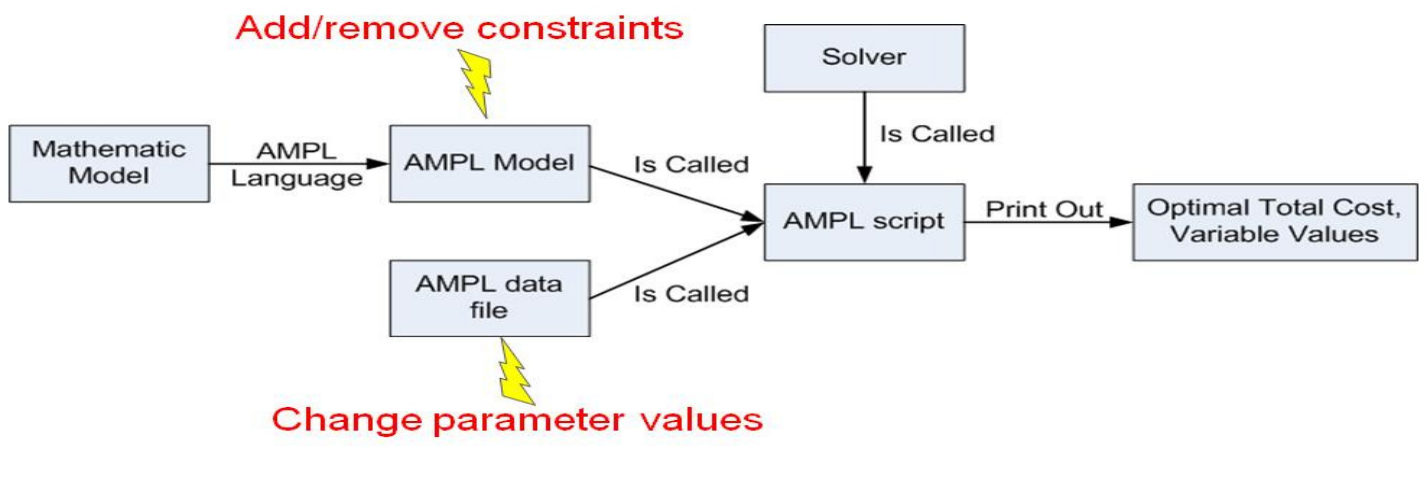

Figure 5. The optimization process

Applying this optimization process, different network topologies and deployment strategies can be considered and evaluated by changing the model (add/remove constraints) or values of parameters accordingly.

### 4.5. Input Data for Parameters

In this paper, we have used the following unit prices for the cost-based model. The server operation cost is referenced from Amazon EC2 [16]. It takes $P_a$ = 10 cents/hour for one VM running on Amazon EC2. Generally it's cheaper to run one VM on the operator's own infrastructure than on a public cloud. Moreover, the secondary sites are smaller and further away (from the centralized management framework) than primary sites, so the server operation cost in primary sites is cheaper than that in secondary sites. Considering these factors, we assign a server operation cost in secondary sites ($P_s$) of 8 cents/hour and operation cost in primary sites ($P_p$) of 6 cents/hour.

The link cost is also referenced from Amazon EC2 [16]. The EC2's price for outgoing data transfer is 12 cents/GB (in average because they apply the accumulative price model). With the average streaming bit rate of *300Kb/s = 0.126 GB/h*, we can calculate the link cost per hour: $L_a = 12 \times 0{,}126 = 1{,}512$ (cents/hour). Again, it is of course cheaper if operators use their own link (back bone) instead of external connections, therefore the link cost between a primary site and a secondary site ($L_s$) is given as 10 cents/GB, or 1,26 cents/hour. Similarly, the link cost between a primary site and another primary site ($L_{pp}$) is chosen as 1.0 cent/hour.

In addition to unit prices, we choose the server capacity as 300, meaning one streaming server can support 300 streaming requests at the same time. Besides, to see how the system handles the peak load using a public cloud, we also decide that the primary site's capacity is 800 servers (medium-sized data center) and the secondary site's capacity is 100 servers (small-sized data center). Table 1 summarizes all the parameters mentioned above.

Table 1. Parameter summary.

| Parameter | Notation | Value | Unit |
|---|---|---|---|
| Operation cost for one server in Primary site | $P_p$ | 6 | cents/hour |
| Operation cost for one server in Secondary site | $P_s$ | 8 | cents/hour |
| Operation price for one server in Amazon site | $P_a$ | 10 | cents/hour |
| Link cost per client between Pri. site and Pri. site | $L_{pp}$ | 1,0 | cents/hour |
| Link cost per client between Pri. site and Sec. site | $L_s$ | 1,26 | cents/hour |
| Link cost per client between Pri. site and Amazon | $L_a$ | 1,512 | cents/hour |
| Streaming server's capacity | U | 300 | clients/server |
| Primary site's capacity | | 800 | servers |
| Secondary site's capacity | | 100 | servers |

## 5. EVALUATION

The cost-based mathematic model in Section 4.3 can be used to analyze and evaluate different aspects of the cloud-based telecom network including the applications described in Section 4.1. We aim to do the following comparisons and evaluations to illustrate that:

1. Compare the total cost of deployment in different network topologies.
2. With a specific network topology, compare the total cost of different deployment strategies.
3. With each network topology, characterize the deployment solution that yields minimal total cost over the entire network.

## 5.1. Comparing Network Topologies

We can evaluate the total cost for network topologies that have different numbers of primary sites or secondary sites by changing the input values for accordant parameters. We can also compare the total cost of different network topologies that have the same number of sites but different in connections between sites by changing the set of assumptions, i.e. remove current assumption and/or add new assumptions. For example, by adding one additional assumption (to the current assumptions in Section 4.1) to get a new network topology:

**Assumption 4**: Previously, primary sites were connected to each other to form a full-mesh network (Figure 6a). However, now we assume that there is no connection between primary sites. This could be the situation when the service spawns to different mobile networks in one or several countries. The new network topology looks like Figure 6b. To implement this new assumption into the model in Section 4.3, we can simply add one more constraint: $F_{pp}[i,j] = 0$. This means the number of redirections between the $i^{th}$ primary site and the $j^{th}$ primary site is 0.

Between the old and new network topologies, operators would want to see how the redirections between primary sites affect the total cost. To do that, we used two network topologies with a subscriber distribution as illustrated in Figure 6. These topologies include 3 branches; each branch consists of 1 primary site and 2 secondary sites. Subscribers connect to branch 1 (50%) and branch 3 (50%), but there isn't any subscriber connecting to branch 2. This subscriber distribution is chosen to simplify the modelling of “free” capacity at branch 2. Available free capacity is the condition to redirect client requests from branch 1 and 3 to branch 2.

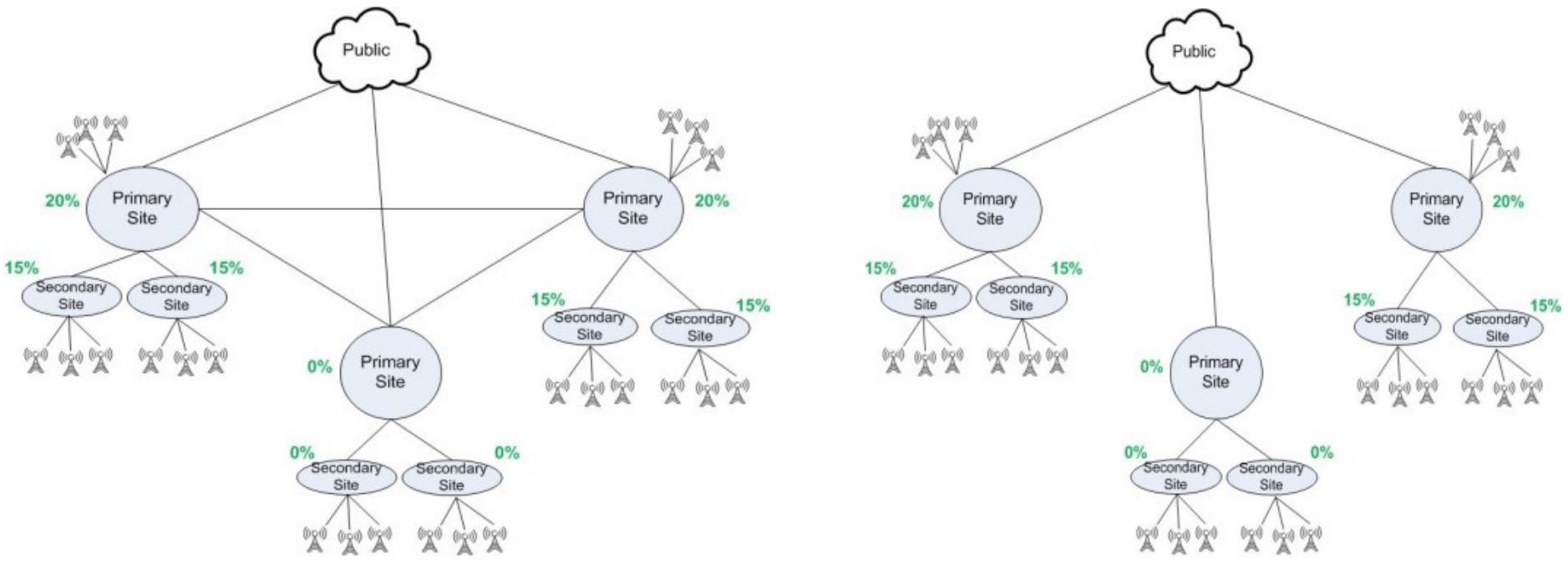

(a) With full-mesh primary site (b) Without full-mesh primary site

Figure 6. Different network topologies

To get the total number of subscribers, we have considered the 2013 data traffic forecasts for a typical medium sized Western European mobile operator in which there are 7 million mobile data subscribers [15]. This number is just an example and it doesn't play any important role to meaning of the result. With the subscriber distribution in the Figure 6, the number of clients connected to each site can be calculated easily.

Running the optimization solving process in Section 4.4, we can find out the minimal total cost (USD/hour) and the number of servers at each site ($n_p$ at primary sites, $n_s$ at secondary sites and $n_a$ public cloud) as well as the number of requests served at each site. The total cost and the number of servers at each site are presented in Table 2.

Table 2. Compare total cost

| | **Topology 6a** | **Topology 6b** |
|---|---|---|
| Total Cost (USD/h) | 145989,40 | 147221,40 |
| $n_p[1]$ (servers) | 800 | 800 |
| $n_p[2]$ (servers) | 800 | 0 |
| $n_p[3]$ (servers) | 800 | 800 |
| $n_s[1,1]$ (servers) | 100 | 100 |
| $n_s[1,2]$ (servers) | 100 | 100 |
| $n_s[2,1]$ (servers) | 0 | 0 |
| $n_s[2,2]$ (servers) | 0 | 0 |
| $n_s[3,1]$ (servers) | 100 | 100 |
| $n_s[3,2]$ (servers) | 100 | 100 |
| $n_a$ (servers) | 20534 | 21334 |

The table shows that the total cost of topology 6a is smaller than the total cost of topology 6b. The reason is that there are redirections between primary sites in topology 6a. When one primary site runs out of capacity, it can redirect its client requests to other primary sites which still has "free" capacity ($n_p[2]$ = 800 servers) instead of redirecting them directly to public clouds. As the price for resource usage in internal primary sites is cheaper than in the public cloud, the topology 6a will generate a cheaper total cost than the total cost of topology 6b.

The current difference between the two total costs is 1232 (USD/hour), and it is equal to ≈ 1% of the total cost of topology 6b. Since the larger parts of the two total costs are the costs for servers on the public cloud (20534 and 21334 servers), this ratio will increase if we choose higher site capacities (the current ones are 800 servers for a primary site and 100 servers for a secondary site) or a smaller total number of subscribers (the current one is 7 million). For example, if capacity of primary sites and secondary sites are 4000 and 3000 servers respectively, then the total cost will reduce 21,41%.

## 5.2. Comparing Deployment Strategies

To compare different deployment strategies, we need to choose one specific mobile network topology as an example. Thus, we re-used the network topology in Figure 6a with a new subscriber distribution (Figure 7) which is more general than the distribution in Figure 6a. Then we aim to answer the following questions:

1. What is the benefit of deploying streaming servers to secondary sites compared to other solutions that deploy streaming servers to primary sites only?
2. What is the economical advantage of deploying streaming servers onto the hybrid cloud compared to other solutions that deploy servers onto internal data center only (without cloud); or deploy all servers onto the public cloud?

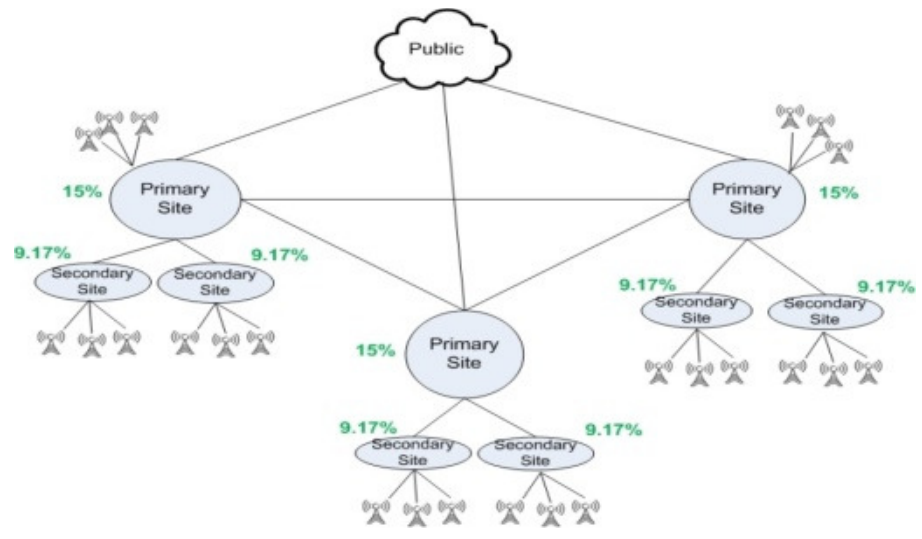

Figure 7. Equal Subscriber Distribution

## A. Deploying streaming server to secondary site

We compared the total costs of two solutions below:

Solution 1: Deploy streaming servers to primary sites only, not to secondary site. All clients connected to a secondary site are redirected and served from a primary site. In this case, the number of redirections is equal to the number of clients connecting to the secondary site (Cs). To express this solution by the model in Section 4.3, we can easily add one more constraint: number of clients served from every secondary site (Ss) = 0.

Solution 2: Deploy streaming servers to both primary sites and secondary sites. Clients connected to a secondary site can be served from that secondary site or be redirected and served from a primary site. In this case, the number of redirections ranges from 0 to the number of clients at the secondary site (Cs).

In this comparison, we also used 7 million for the total number of subscribers as in Section 5.1. With the subscriber distribution in Figure 8, we have 1050000 users at each primary site and 641900 users at each secondary site. These values are put into the data file for number of clients at primary and secondary site parameters (Cp, Cs) together with input values for other parameters as in Section 4.5. After that, we run the optimization solving process in Section 4.4 and get the minimal total costs of the two solutions as in Table 3.

Table 3. Cost Comparison

| Solution | Total Cost |
|---|---|
| S1: Don't deploy to secondary site | 142675,6 |
| S2: Deploy to secondary site | 137803,6 |

The table shows that solution S2 is cheaper than solution S1. The difference is 4872 (USD/hour) and it is equal to ≈ 3.4% the total cost of the solution S1. For the same reason as in Section 5.1, this ratio will increase if we choose higher site capacities or a smaller total number of subscribers. For example, if capacity of primary sites and secondary sites are 8000 and 3000 servers respectively, then the total cost will reduce 96,5%. Thus, although it is not a common practice in today's network to deploy server type nodes, such as caches or media streaming servers, to secondary sites, the mathematic model has figured out that deploying servers to secondary sites is more beneficial than deploying servers to primary sites only.

## B. Dynamically scaling to Public Cloud

We have run the model with the number of clients connecting to primary and secondary sites changed from 0 to the site capacity and more. From Section 4.5 we have: the capacities of primary site and secondary site are 800 and 100 servers respectively. Additionally, one server's capacity is 300 concurrent streams. Therefore, the primary site capacity is: 800 x 300 = 240.000 (clients) and the secondary site capacity is: 100 x 300 = 30.000 (clients).

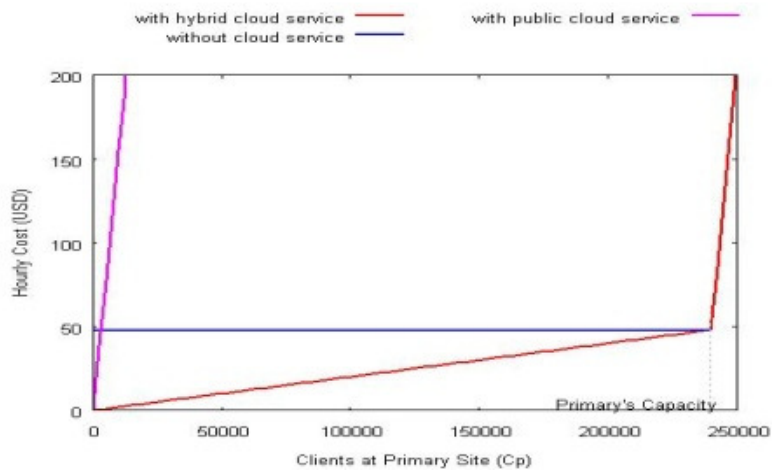

Figure 8. Hourly cost at a primary site

Figure 8 shows the hourly total cost over the number of clients at a primary site. The cost graph of the hybrid cloud service is the red line. When the number of clients is in between 0 and 240000, the total cost increase steadily since the service deploys a sufficient number of servers in the primary site to serve all client requests. As clients are connected and served directly from the primary site, there is no link cost for transferring data, only server operation cost. When the number of clients is bigger than the primary site's capacity (> 240000), the service starts to deploy servers to a public cloud to serve the excess requests. Therefore, the total increases dramatically because there are costs for running servers in the public cloud and transferring data from public cloud to primary site.

The pink line in Figure 8 is the total cost graph if operators deploy streaming servers to public cloud only. It increases rapidly. The blue line is the total cost graph if operators deploy servers to the primary site without cloud technology. The total cost graph in this case is constant all the time because all hardware capability of the primary site is dedicated to the streaming service only, even there is a small number of client requests coming to the site (to deal with peak loads). Moreover, when the number of requests increases over the site capacity, the service will not work.

From Figure 8 it is clearly seen that the deployment solution using the hybrid cloud is cheaper while it still guarantees that services will work all the time compared with the other two solutions: using private primary sites without cloud technology or using a public cloud only.

At secondary sites, the story is quite similar; there is only one difference: if a secondary site runs out of capacity, the requests coming to the secondary site will be redirected to the connected primary site before they are redirected to the public cloud. If there is “left-over” capacity at the primary site, redirected requests would be served from there until the primary site runs out of capacity as well.

## 5.3. Characterizing Optimal Deployment

In order to characterize the optimal deployment for each network topology, we have used the network topology illustrated in Figure 8 as a reference network topology to compare the total cost of two solutions below:

Solution 1: All clients connecting to each primary and secondary site are served from the closest site or the one that they are connected to. There is no redirection. We can use the mathematic model to express this solution by changing the constraint C5 (Section 4.3) to: $C_s[i,j] = S_s[i,j]$. This guarantees that all clients connecting to a secondary site will be served from that secondary site.

Solution 2: The client requests can be redirected - may be not to the closest site - to obtain the minimal total cost for the operator. The number of redirections can be any number between 0 and the number of clients at the secondary site ($C_s$) as long as the total is minimal.

In this comparison, we have changed the number of clients connecting to a secondary site within a range, i.e. 20000-40000, and compare the total costs of the two solutions. The results show a fine-grained optimization. From the results, we can also find out when the redirections happen to reduce the total cost and how much they could save (see [14]} for more details) for the operator. Currently, the saving amount is quite small (i.e. ≈ 0.5\% for the range 20000-40000); however, the computing time of the optimization process is very negligible. Besides the cost decrement itself, this also indicates that locating servers as close to clients as possible is not always the optimal deployment.

In summary, we have used the cost-based mathematic model in Section 4.3 to analyse three different scenarios. Firstly, we compared the optimal total cost of two different telecom network topologies. The result shows a noticeable difference between these two total costs. Secondly, with one reference network topology, we compared the total cost of two different deployment strategies. The result shows a great cost reduction of using the appropriate deployment strategy. Finally, we optimized the total cost of the same reference network topology (used in the second scenario). The result shows a very fine-grained optimization. Thus, the mathematic model proved to be a suitable tool for analysing different aspects of service deployment in Telco clouds from a cost-reduction perspective.

## 6. TECHNOLOGY VERIFICATION

In this section, we will describe the deployment of a prototype to demonstrate the dynamically scaling characteristic of the proposed streaming service: operators could deploy streaming servers dynamically into distributed telecom sites located worldwide as well as public clouds to meet the needs of clients (see [21] for more details). The prototype is a complete system that provides a live streaming service to the end-user. Its architecture is presented in Figure 9. It consists of two parts: the local data center and the public cloud Amazon EC2 [16]. In the local data center, we utilize the existing infrastructure including the cloud manager which can allocate resources in hybrid clouds [7] and two private clouds: Eucalyptus [18] and VMware [17].

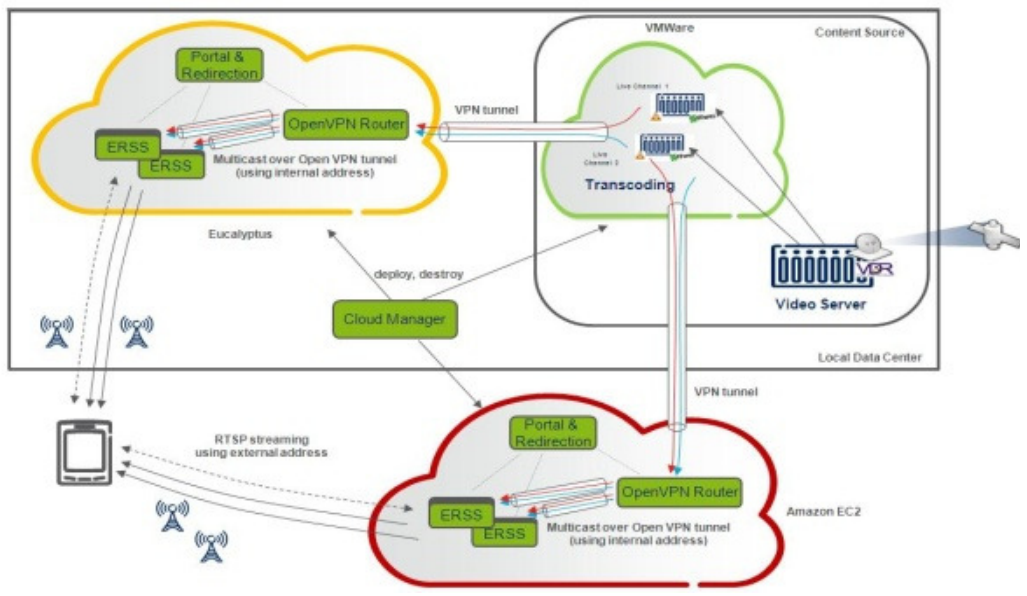

Figure 9. The architecture of streaming prototype [21]

### 6.1. Live Content Source

The content source of the prototype is a live broadcast TV program fetched from a satellite signal. This signal is recorded and transcoded into transport streams (mpegts) which is the required input of the Ericsson streaming server introduced in the next section. In order to transcode the mpeg file, we use the FFmpeg [20] transcoder with the X264 library. The transcoder is deployed in the VMWare cloud for the purpose of easy scalability and multiple format support capability. For example, in Figure 9, there are two encoders to encode two different live channels. We can even deploy more encoders to increase encoding capacity or support more channels.

### 6.2. Streaming Server Nodes

The streaming server node is a virtual machine with the installed Ericsson Research Streaming Server (ERSS). It is created automatically by the cloud manager through two steps: first, the cloud manager starts an empty VM, and then it deploys the streaming server into that VM. This node receives a transport stream (described by a configuration file) from content source as the input and provides a RTSP stream as the output.

To demonstrate the dynamically scaling characteristic of the system based on the demand, the cloud manager will dynamically deploy streaming servers onto the private cloud Eucalyptus or the public cloud Amazon EC2. The more requests come, the more streaming servers are deployed.

### 6.3. OpenVPN Server Nodes

We have considered two noticeable points in sending the live content stream from the FFmpeg transcoder to the streaming servers:

- First, because of the timely delivery and package-loss-tolerance requirements of real-time multimedia streaming applications, we choose to use a VPN tunnel to send the content stream from FFmpeg to ERSS instead of plain TCP/IP.
- Second, if we send the stream from FFmpeg directly to each ERSS server following the unicast model, it could cause a bottle neck at the FFmpeg encoder and also cost more money for transferring data into Eucalyptus and EC2 clouds (where the ERSS servers are placed).

Addressing these two points, we have applied the idea of the multicast model by introducing an intermediate OpenVPN server node as in Figure 9. This node acts as a router for all streaming servers connected to it (on the Eucalyptus cloud or EC2 cloud). It receives the stream from the transcoder through the OpenVPN tunnel and then broadcasts the received stream to all connected Streaming Server (ERSS) nodes also through the OpenVPN tunnels. This model helps to reduce the outgoing traffic at the transcoder and also reduce the data transfer in/out of Amazon EC2, thereby reducing the cost.

### 6.4. Demonstration and Analysis

We have run a demonstration in which we deployed one OpenVPN server, one FFmpeg encoder, and one streaming server ERSS. It took only 7 - 9 minutes to deploy all of these three nodes, and 2-3 minutes for each node. After that, users can start playing streaming from mobile or PC.

To demonstrate the scaling capability of the service, we assumed that the number of requests increases, so we deploy several streaming servers ERSS more to serve new requests. To deploy an ERSS server, it took only 3 minutes. Additionally, assuming the number of requests goes down, we destroyed some streaming server. To destroy a server, it took only 1 minute.

To illustrate how the service could move the peak load onto public clouds, we assumed that the Telco Cloud has been run out of capacity while new requests are still coming. In this case, operators need to rent external resources to serve these new requests; otherwise the service will not work. Thus, instead of deploying the OpenVPN server and streaming servers onto the private cloud Eucalyptus, we deployed them onto the public cloud Amazon EC2. The deployment was performed successfully within similar time as above (2-3 minutes/server).

In short, the prototype has proved that operators completely could deploy streaming servers dynamically into worldwide telecom sites (primary and secondary) as well as public clouds as described in section 3 and section 4. As a proof-of-concept, it has successfully resolved related technical issues to make the service work and thereby provide a practical evidence for the cloud service and the model.

## 7. CONCLUSION

We have proposed an applicable streaming service architecture with necessary components and working mechanism. This architecture sharpens the new business opportunity for operator in offering competitive cloud services. Furthermore, we have modelled that architecture as a cost based mathematic model to analyse several aspects of a mobile network. First, we compared two different network topologies. Our model showed that a topology with full-meshed primary sites has a slight cost advantage ≈ 1% over a topology with no inter-primary-site links. Taking the full-meshed topology, we analysed deployment of media servers in primary and secondary sites vs. deployment in primary sites only. The result suggests that using secondary sites is more cost efficient ≈ 3.4%. Finally we analysed a quite fine-grained optimization that takes advantage of virtualization technology, showing a cost reduction of up to 0.5% depending on number and distribution of clients. All these results depend of course on the input parameters, such as cost for internal links. Each network provider will have its own numbers and so the results will look slightly different. In short, our approach of modelling the mobile network and its cloud-related properties mathematically proved to be a quick and efficient way to analyse relevant aspects of mobile network configurations from a cost-perspective. This result is to be seen as a foundation and framework for future research in this area. Since the actual calculation time is negligible, this approach could e.g. be used for real time monitoring and managing the "Telco Cloud"', in order to exploit its dynamic properties in an optimal way. The feasibility of dynamic server deployments for a multimedia application in a cloud environment was demonstrated in the lab at Ericsson Eurolab.

Our optimization framework can be extended in various ways. Parameters such as "number of clients ($C_p$, $C_s$)" could be changed to variables to well reflect the high fluctuations in network requests. This would make the objective function nonlinear, but with our mathematic model, we can just choose a suitable nonlinear solver. Another way, by understanding the traffic patterns, operators can apply different workload prediction strategy to let cloud manager know the anticipated demand and then scale application up and down with proper planning. Thus, the service can be scaled pro-actively, instead of reactively as the moment. We could also develop an useful tool with nice interface allowing operators to add or remove parameters and variables, formulate a total cost and constraints, and then generate different AMPL models and data files according to their needs. An example for that customizing demand is that operators want to roll out the model for many countries and they would need a so-called "master primary site"'. In this case, the model would change and the problem would be nonlinear as well.